\documentclass[runningheads]{llncs}
\usepackage{graphicx}

\usepackage{comment}

\usepackage[shortlabels]{enumitem}
\newcommand{\bit}{\begin{itemize}}
\newcommand{\eit}{\end{itemize}}

\usepackage{amsfonts}
\usepackage{multirow}
\usepackage{latexsym}
\usepackage{epsfig}
\usepackage{epstopdf}

\usepackage{bm}

\usepackage{color}
\usepackage{algorithm, algorithmic}

\usepackage{amsmath, amssymb} 
\usepackage{graphicx}
\usepackage[shortlabels]{enumitem}
\usepackage{thmtools}

\newcommand{\rr}{\textcolor{red}}
\newcommand{\bb}{\textcolor{blue}}

\usepackage{calc}

\begin{document}
\title{Transparency: Motivations and Challenges\thanks{Supported by The Alan Turing Institute, Darwin College and the Leverhulme Trust.}}
%
%\titlerunning{Abbreviated paper title}
% If the paper title is too long for the running head, you can set
% an abbreviated paper title here
%
\author{Adrian Weller\inst{1,2,3}\orcidID{0000-0003-1915-7158}}
\authorrunning{A. Weller}
% First names are abbreviated in the running head.
% If there are more than two authors, 'et al.' is used.
%
\institute{University of Cambridge, UK,  \email{adrian.weller@eng.cam.ac.uk}\\
\and
The Alan Turing Institute, UK
 \and
Leverhulme Centre for the Future of Intelligence, UK
}
\maketitle              % typeset the header of the contribution

\begin{abstract} 
Transparency is often deemed 
critical to 
enable effective 
real-world deployment of intelligent systems. 
Yet the motivations for and 
benefits of different types of transparency can vary significantly depending on context, and objective measurement criteria are difficult to identify. 
We provide a brief survey, suggesting challenges and related concerns, particularly when agents have misaligned interests. 

We highlight and review
settings where transparency may 
cause harm,  
discussing connections 
across privacy, multi-agent game theory, economics, fairness and trust. 

\end{abstract}

\section{Introduction}

The case for transparency has been made in many settings, including for government policy \cite{VisKau11}, business \cite{Low96}, charity \cite{SriBat08}, and
algorithms \cite{Mor14}. 
Within machine learning, there is  
a 
general feeling that ``transparency'' -- like ``fairness'' -- is important and good. Yet both concepts are somewhat ambiguous, and can mean different things to different people in different contexts. We 
discuss 
various types of transparency in the context of human interpretation of algorithms, noting 
their benefits, motivations, difficulties for measurement, and potential concerns. 

We then consider settings where, perhaps surprisingly, transparency may lead to a worse outcome.  
Transparency is often beneficial but  
it is not a universal good. We  
draw attention to work in other disciplines and hope to contribute to 
an exploration of which types of transparency are helpful to whom in which contexts,  
while recognizing when conditions may arise such that transparency could be unhelpful. 

We summarize our main themes: \\
(A) There are many types of transparency with different motivations -- we need better ways to articulate them precisely, and to measure them (Section \ref{sec:types}). \\
(B) We should recognize that sometimes transparency is a means to an end, not a goal in itself (Section \ref{sec:types} and Section \ref{sec:means}). \\
(C)  
Actors with misaligned interests can abuse transparency as a manipulation channel, or inappropriately use information gained  
(Section \ref{sec:harms}).\\ 
(D) In some settings, more transparency 
can lead to less efficiency (Section \ref{sec:braess} reviews 
related work in economics, multi-agent game theory and network routing), fairness (Section \ref{sec:fair}) or trust (Section \ref{sec:govt} and Section \ref{sec:trust}).

In Section \ref{sec:machine}, we 
note `machine interpretability' as an important research direction, which may also provide insight into how to measure human understanding in some settings.

\paragraph{Related Work}

There is a considerable literature on transparency and social good. Much of this focuses on the benefits of transparency but some earlier work, notably in economics and social science, also considers drawbacks of transparency and accountability \cite{LerTet99,Prat05wrong,Etz10,Pel11,Ber14}. 
We discuss related work throughout the text.

\section{Types of Transparency: Benefits, Measurement and Motivations}\label{sec:types} 

Considering transparency of algorithmic systems broadly, there are important areas to consider beyond just the algorithm. For machine learning systems trained on data, knowing where and how the data was gathered can be critical, along with understanding who made those choices and what their motivations were. Further, we should look at the socio-technical context of a system to understand how it will be used in practice.

We briefly describe various types of transparency in the context of human interpretability of algorithmic systems, highlighting different possible motivations. We typically seek an explanation  
in understandable terms, which can often be framed as answering questions of ``what'', ``how'', or ``why''  
(either \emph{toward what purpose} in the future, or \emph{due to what cause} in the past).  
Some of our observations have been made previously \cite{Lip16,DosKim17,ijcai}. 
To our knowledge, in the 
setting of artificial intelligence, we make new points on motivations and on measuring understanding.

An automated explanation might arise immediately from the original system -- typically if it has been constrained to lie in some set of classifiers deemed to be interpretable (e.g. a short decision list). Alternatively, a second explainer algorithm may have produced an explanation for the original system. 
We consider various classes of people: a \emph{developer} is building the system; a \emph{deployer} owns it and releases it to the public or some user group; a \emph{user} is a typical user of the system. 
For example, developers might be hired to build a personalized recommendation system to buy products, which Amazon then deploys, to be used by a typical member of the public. People might be experts or not. 

We list several types and goals of transparency. Each may require a different sort of explanation, requiring different measures of efficacy:
%\begin{enumerate}[leftmargin=*, label={Type} \arabic*:]
\begin{description}
\item[Type 1] For a developer, to understand how their system is working, aiming to debug or improve it: to see what is working well or badly, and get a sense for why.
\item[Type 2] For a user, to provide a sense for what the system is doing and why, to enable prediction of what it might do in unforeseen circumstances and build a sense of trust in the technology.
\item[Type 3] For society broadly to understand and become comfortable with the strengths and limitations of the system, overcoming a reasonable fear of the unknown.
\item[Type 4] For a user to understand why one particular prediction or decision was reached, to allow a check that the system worked appropriately and to enable meaningful challenge (e.g. credit approval or criminal sentencing).
\item[Type 5] To provide an expert (perhaps a regulator) the ability to audit a prediction or decision trail in detail, particularly if something goes wrong (e.g. a crash by an autonomous car). This may require storing key data streams and tracing through each logical step,  
and will facilitate assignment of accountability and legal liability.
\item[Type 6] To facilitate monitoring and testing for safety standards. 

\item[Type 7] To make a user (the audience) feel comfortable with a prediction or decision so that they keep using the system. 
Beneficiary: deployer. 
\item[Type 8] To lead a user (the audience) into some action or behavior -- e.g. Amazon might recommend a product, providing an explanation in order that you will then click through to make a purchase. Beneficiary: deployer.
\end{description}
%\end{enumerate}

We can differentiate between the intended \emph{audience} of an explanation and the likely \emph{beneficiary} (or beneficiaries). We suggest that types  
1-6 are broadly beneficial for society provided that explanations given are \emph{faithful}, in the sense that they accurately convey a true understanding without hiding important details. 
This notion of faithful can be hard to characterize precisely. It is similar in spirit to the instructions sometimes given in courts to tell ``the truth, the whole truth, and nothing but the truth.''

Defining criteria and tests for practical faithfulness are important open problems. We suggest that helpful progress may be made in future 
by focusing on one particular context at a time. We make a similar suggestion for the challenges of characterizing if an explanation is good at conveying faithful information in understandable form, and if a human has actually understood it well.\footnote{Greater faithfulness of an explanation may challenge the ability of its audience to understand it well, perhaps requiring a greater investment of time and effort 
\cite{LIME}.}  
\cite{DosKim17} suggest several methods, such as establishing a quantitative approximate measure (e.g. if we are restricting models to be decision trees then we can feel reasonably confident that a model becomes harder to understand as the number of nodes increases), or asking a human if they can correctly estimate what the system would output for given inputs. We suggest further approaches below and highlight a key challenge.

We use (as in \cite{DosKim17}) the terms \emph{global} interpretability or explanation for a general understanding of how an overall system works, as in our transparency types 2-3; and \emph{local}  interpretability for an explanation of a particular prediction or decision, as in types 4, 5, 7 and 8 (though both forms may be useful for a given type). 

For global interpretability, we mention two interesting possible approaches due to quotes attributed to the physicist Richard Feynman: (i) ``What I cannot create, I do not understand'' suggests that in some settings, a good test of understanding might be to see if the person could recreate the whole system (given expert help and allowing some reasonable tolerance); 
(ii) ``If you can't explain it to a six year old, you don't really understand it'' suggests a possible  
meta-approach to test clarity of an explanation -- for any given  
test $T$ of human understanding,  
ask the person 
to explain the system to someone new, then give that new person the test $T$. 
\cite{Lak16} 
introduced measures of human interpretability based on being able to describe a decision boundary, which would facilitate model reconstruction.   
Rudin \cite{Rud18} argues that in many settings, a global model which is constrained to be `interpretable by construction', for example by being described by a simple, short, intuitive scoring system, will provide sufficiently high performance \cite{RudUst18}. 
We discuss further  
notions of understanding in Section \ref{sec:machine}.

\begin{figure}[t]
\centering
\includegraphics[width=0.4\linewidth]{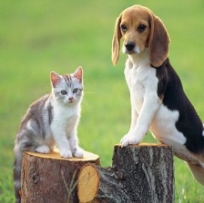} \quad
\includegraphics[width=0.4\linewidth]{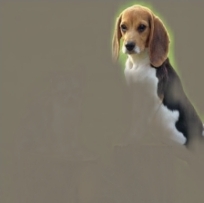}
\caption{\small An image (left) is given to a classification system. A separate explaining algorithm extracts the sub-image (right) which it estimates led the original system to output ``beagle'' (from \cite{DabGal17}). Would it be better or not if the sub-image contained the legs of the dog? That depends on the classification system, and what exactly is desired, and is not simple to answer, demonstrating one challenge in defining a quantitative measure of the quality of the explanation.}
\label{fig:salient}
\end{figure}

Highlighting a key challenge even for local interpretability, consider explaining the output of an image classification system, as illustrated in Figure \ref{fig:salient}. 
Several recent approaches attempt to identify the parts of a given image which are most salient, i.e. those parts which in a sense were most responsible for leading to the system's prediction \cite{LIME,Zin17}. Such approaches can be very helpful -- e.g. for type 1 transparency, we might learn that a system which reliably tells apart wolves from huskies on a test set might in fact be relying on the presence or absence of snow in the background, rather than features of the animal itself, and hence may be unlikely to generalize well on test data \cite{LIME}. 
Suppose we are given two such methods of generating a salient sub-image for a classification system. How should we measure which method provides a better explanation of what the system is doing? This is an important question, where the answer will depend on sharpening our understanding of exactly what we are seeking. Note that it is not helpful to compare against what a human thinks is relevant in the image. %We should like
Rather, we want a sub-image which is high in predictive information \emph{for the system} yet is not too large, focusing only on the relevant region. A promising possible direction for a quantitative solution was suggested by \cite{DabGal17}, who propose a measure of %\emph{
concentration of information. 

\cite{Sam17} proposed a different quantitative metric for evaluating methods which return ordered collections of input pixels, based on region perturbation. On this metric, an approach called Layer-wise Relevance Propagation (LRP \cite{Bach15}) performed well for explaining classifications made by deep neural networks. 
Alternative `axiomatic' approaches have been developed for identifying and quantifying the contributions of each input feature toward a particular classification, based on specifying reasonable desirable properties which an explanation should have \cite{Strum14,Datta16,Sun17}. Interestingly, these approaches link to earlier work by Shapley \cite{Shap53} on determining the value of contributions made in $n-$person games. Typically, it is computationally intractable to compute the appropriate contributions exactly, but various methods may be regarded as approximations to this approach \cite{Lun17}.

We briefly note two other approaches to \textit{ex post} explanations of a specific automated classification. Rather than provide an attribution over features of the input, \cite{Koh17} instead identifies which training data points were particularly influential for making the classification. 
\cite{Wac18} propose \emph{counterfactual explanations}: suppose an individual applies to a bank for a loan but is classified as not being sufficiently creditworthy; a counterfactual explanation reveals the minimum change required in some feature(s), for example income, such that the loan would instead be classified as approved, thus potentially providing an action which could feasibly be taken by the individual to change the decision. 

Other approaches seek to identify interpretable representations in order to help understand how an algorithm works~\cite{infoGAN,betaVAE,Adel18}. We have not provided an exhaustive survey of interpretability methods, but hope that it is clear that different approaches yield different notions of transparency, each of which may be useful in different settings. There is no universally appropriate approach. 

\paragraph{Transparency as a Proxy}
Transparency can %serve as a proxy for 
provide insight into other characteristics which may be %even 
hard to measure (as noted by \cite{DosKim17}). We noted above how local explanations for an image classification system -- revealing how wolves were differentiated from huskies -- demonstrated the lack of robustness of the system. Other features where transparency can provide helpful insight include safety, fairness, verification and causality. 
There is a rapidly growing literature on methods to try to address these areas directly \cite{Dwork12,feldman_kdd15,hardt_nips16,joseph_bandits,Amo16,VarAle16,ijcai,PetJanSch17}.

\section{Possible Dangers of Transparency}\label{sec:harms}

In this section, 
we begin to examine ways that transparency may be unhelpful.

\subsection{Divergence between Audience and Beneficiary}

There are some forms of transparency, such as types 7 and 8 in Section \ref{sec:types}, where the intended audience for an explanation diverges from the beneficiary, hence the motivation may be suspect. This can lead to worrying types of manipulation and requires careful consideration. 

Considering type 7 transparency, we draw attention to the remarkable 
`Copy Machine' study. \cite{Lan78} arranged for researchers to try to jump in line to make a few photocopies at a busy library copy machine. The researcher either (i) gave no explanation, asking simply ``May I use the xerox machine?''; (ii) provided an `empty' explanation:  ``May I use the xerox machine, because I have to make copies?''; or (iii) provided a `real' explanation: ``May I use the xerox machine, because I'm in a rush?'' The respective 
success rates 
were: (i) $60\%$; (ii) $93\%$; and (iii) $94\%$. The startling conclusion was that saying ``\emph{because} something'' seemed to work  
%very 
effectively to attain compliance, even if the `something' had zero information content. 
Hence, a possible worry is that a deployer might provide an empty explanation as a psychological tool to soothe users.

In fact there is a line of research which considers all communication 
often to be more a form of manipulation than a way to transfer information. This view is 
prominent when taking an evolutionary view of multiple agents \cite{Wil83}, a perspective which we revisit throughout the remainder of this paper. 
Earlier work by \cite{Adel79} explored whether disclosures provided in financial reports are more aptly described as communication or manipulation.

For type 8 transparency, where the deployer has a clear motive which may not be in the best interests of the audience of the explanation, particular care and future study is warranted. 
Even if a faithful explanation is given, it may have been carefully selected from a large set of possible faithful explanations %, carefully chosen 
in order to serve the deployer's goals.

\subsection{Government Use of Algorithms}\label{sec:govt}
In many states in the US, a private company provides the COMPAS system to judges to help predict the recidivism risk of a prisoner, i.e. the chance that the prisoner will commit a crime again if released. This information is an important factor in parole hearings to determine whether to release prisoners early or to keep them locked behind bars.  
Significant attention has focused on whether or not the prediction system is fair \cite{Ang16}. We discuss  connections between fairness and transparency in Section \ref{sec:selective} and Section \ref{sec:fair}. Here we consider the appropriate degree of transparency of such a system: a prisoner should at least have transparency type 4 from Section \ref{sec:types} in order to check if proper process has been followed and enable potential challenge, but can there be too much transparency?

Perhaps motivated by concerns in the US over the COMPAS system, Bulgaria passed legislation requiring that (many forms of) government software be open source, ``after all, it's paid by tax-payers' money and they should... be able to see it'' \cite{Cold16}. 
More recently, New York announced the creation of a task force to examine how city agencies use algorithms to make decisions, looking to find ways to make automated systems more transparent, fair and accountable \cite{Wig18}. 
%We should clarify 
Note that a machine learning system typically consists of both algorithms and data, and having access to just one of these may not provide much meaningful information. %be very useful. 
We consider the %extreme 
case of all algorithms and data being transparently available and
note several concerns.

\paragraph{Gaming and IP Incentives}
If all details are readily available, this can facilitate gaming of the rules \cite{Ghani16}.\footnote{One view is that if rules are set up correctly, then transparency will not lead to `gaming' %is impossible 
since agents optimizing their own objectives subject to the rules will necessarily lead to a good outcome for all. However, it is often very challenging in practice to get the rules exactly right in this way -- thus there may be a distinction between the `letter' and the `spirit' of the law. 
See Section \ref{sec:braess} for a related example. 
}  
In addition, if all code and data is open source, then this reduces incentives to develop relevant private intellectual property, which may delay progress significantly. 

\paragraph{Transparency and Privacy}
Indeed, in many cases, transparency may be viewed as the opposite of privacy. Many in society feel that some sort of right to privacy -- and hence, a limit to transparency -- is appropriate. 
Legal frameworks vary by country. A recent landmark decision by the Supreme Court of India is explicit, stating ``The right to privacy is protected as an intrinsic part of the right to life and personal liberty under Article 21 and as a part of the freedoms guaranteed by Part III of the Constitution'' \cite{India17}. 
Tensions between privacy and transparency can exist even for one user whose data is used in a system -- the user may want their personal data to be kept private but might also like a right to an explanation of how that same system (algorithms + data) works.  
 
Further, 
there are many settings where privacy (i.e. a lack of transparency to all) is critical to foster a trusting relationship of confidence. Examples include the relationship between a doctor and a patient, a lawyer and their client, or discussions of international diplomacy. Inside these relationships, it is interesting to question whether greater transparency leads to trust. We return to this topic in Section \ref{sec:trust}. Here we %consider the opposite direction, and 
rather suggest that a prudent approach is to release private information only to a partner that is already trusted, hence trust can lead to transparency. Providing information to an agent empowers them \cite{Sol01}, hence you should first be confident that their interests align with yours.

As one example, there has been discussion about the extent to which government agencies such as the NSA should be allowed to collect data on individuals. Some argue that ``if you've got nothing to hide, you've got nothing to fear'' \cite{Sol11} or suggest that collecting only `metadata' is harmless. Yet \cite{Cole14} quotes General Michael Hayden, former director of the NSA and CIA, as saying ``We kill people based on metadata.''

\subsection{Means and Ends}\label{sec:means} 
In Section \ref{sec:types}, we noted that transparency can serve as an imperfect proxy to gain insight into other desirable properties of a system, such as reliability or fairness. 
For example, transparency is often cited as critical for deployment of autonomous vehicles. We suggest that these transparency concerns are primarily for types 1, 2, 3, 5 and 6. Each of these involves somewhat different types of explanation or understanding, with some easier to implement than others. A key concern is reliable safety: how can we be certain that the vehicle will perform well in all circumstances if we do not understand exactly how it is working?

In such cases, we should take care not to stifle innovation by confusing transparency as an end in itself rather than a means to other goals. 
It is conceivable that much time and resources could be spent trying to gain an extremely transparent understanding, when those efforts might be better spent directly on the goal of improving safety. Being pragmatic, it is plausible that society will resort to implementing various safety tests, such as are used for aircraft autopilot systems. If such tests are passed by an autonomous vehicle system, and accidents are extremely rare in practice, then it may make sense to proceed even without full transparency. After all, which is preferable: full transparency and many deaths per year from accidents, or %much 
less transparency and %far 
fewer deaths per year? As \cite{Ein41} eloquently put it, 
%said, 
we must beware the ``perfection of means and confusion of goals.''

\section{Economics and Multi-Agent Game Theory}\label{sec:braess}

In an economy, each individual can act as an autonomous agent. If each agent optimizes her own selfish utility, there is no guarantee in general that this will lead to the best outcome for society. Under restrictive assumptions, \cite{ArrDeb54} famously proved the existence of a general equilibrium for a competitive economy. 
If \emph{externalities} are present, i.e. if costs of one agent's actions fall on others, then we should expect that the result of each agent optimizing her own outcome may lead to a suboptimal result for the whole. This phenomenon is sometimes described as the \emph{price of anarchy} \cite{Rough06}.

We relate this to transparency by considering what happens if all agents are given additional faithful information about a system. 
This  
is a form of type 3 transparency from Section \ref{sec:types}, which may seem the most innocuous of all. %}   
An engineering perspective might naturally lead one to suspect that more information should lead to a better outcome. However, the background of an economist or multi-agent game theorist helps to realize that more information empowers the agents to optimize their own agendas more efficiently, and thus may lead to a worse global outcome. We illustrate these ideas with a striking example of Braess' paradox \cite{Bra68} as given by \cite{Kel91}.

\begin{figure}[t]
		\begin{center}
		\includegraphics[width=0.95\textwidth]{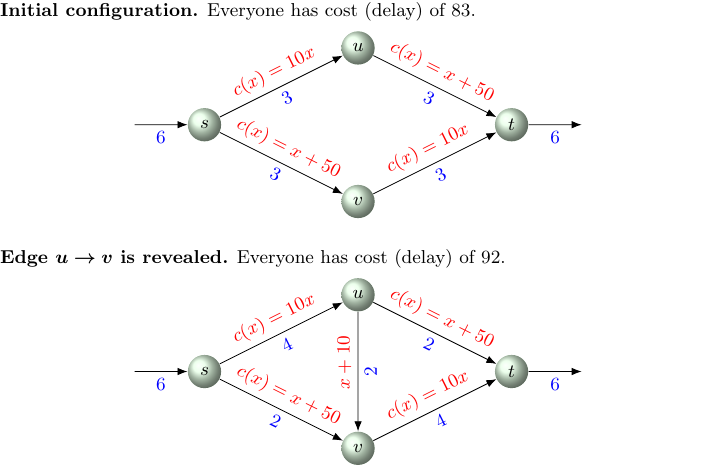}
		\end{center}
		\vspace{-0.7em}
		\caption{\small An example of Braess' paradox for network flow from \cite{Kel91}. 6 cars of flow must pass from $s$ to $t$. For each edge: \rr{red} (above) shows the \rr{cost} (i.e. delay incurred) as a function of the flow $x$ through it; \bb{blue} (below) shows the \bb{Wardrop equilibrium flow} given the costs, based on each car selfishly optimizing. When the edge $u \rightarrow v$ is revealed, surprisingly everyone does worse.}
		\label{fig:Braess}
\end{figure}

Figure \ref{fig:Braess} (top) shows a traffic network where 6 cars enter $s$ at the left, flow through the network via either $u$ or $v$ and exit from $t$ at the right. The costs (delays) of each edge are shown in red, and importantly rise as the amount of flow through them increases. This is realistic in that greater traffic flow on a road often leads to longer delays for everyone on it. Thus we have externalities. If each agent optimizes her own utility, the \emph{Wardrop equilibrium} shown in blue is reached, where each car incurs a delay of $10 \cdot 3 + 3 + 50=83$ time units (details in \cite{Kel91}; 
at a Wardrop equilibrium, no individual can reduce her path cost by switching routes, hence all routes have the same cost).

Now consider what happens if all cars learn about an extra road from $u$ to $v$ as shown in Figure \ref{fig:Braess} (bottom). We may assume that the $u \rightarrow v$ road was always there but that it was hidden until the faithful information about its existence was made transparent.  
Some cars on the $s \rightarrow u$ path save time by taking the new $u \rightarrow v$ road then $v \rightarrow t$, rather than going directly along $u \rightarrow t$. Cars on $s \rightarrow v$ see an opportunity to reduce their delay by switching to this new $s \rightarrow u \rightarrow v$ route.  
Although intuitively the additional road increases capacity for all and hence seemingly should only lead to a better outcome, in fact the new Wardrop equilibrium obtained by selfish optimization results in a greater delay for everyone of $10 \cdot 4 + 2 + 50 = 92$ units!

\cite{Kel91} provides an additional twist on this example. Suppose now that all users know that road works might be under way on the $u \rightarrow v$ road, which has delay $x+10+R$, where $R$ is a random variable taking the values 0 or 30 with equal probability. If $R$ is unknown, then the expected delay along $u \rightarrow v$ is $x+25$, leading to the outcome in Figure \ref{fig:Braess} (top) with delay of 83. On the other hand, if instead $R$ is known to everyone, then: we either have $R=30$ which leads to the outcome in Figure \ref{fig:Braess} (top); or we have $R=0$ which leads to the outcome in Figure \ref{fig:Braess} (bottom); for an expected delay of $\frac{1}{2}(83+92)=87.5 > 83$. Hence, again transparency (providing everyone with faithful information) leads to a worse expected outcome for everyone.

\subsection{Selective Transparency, Fairness and Policy}\label{sec:selective}

The examples above show that providing full transparency of information to everyone can sometimes result in a worse outcome for all. But what if transparency is provided selectively only to \emph{some} participants? 

Consider the original example of Braess' paradox in Figure \ref{fig:Braess}. Suppose the population is divided into a small privileged subgroup $P$ of size $\epsilon$, and everyone else $Q$. It is not hard to see that if the information about %the presence of 
the $u \rightarrow v$ road is made available only to $P$, then $P$ will do (significantly) better while $Q$ will do (slightly) worse. But now consider if the flow cost functions for $s \rightarrow u$ and $v \rightarrow t$ are slightly changed to be locally flat for flows just above 3, before resuming their increase.\footnote{ 
Consider 
$c(x)= \begin{cases}
10x & x \leq 3\\
30 & 3 \leq x \leq 3+\epsilon\\
10(x-\epsilon) & 3+\epsilon \leq x.
\end{cases}$}   
Now \emph{everyone does better} if the $u \rightarrow v$ road is revealed \emph{only to} $P$ (though $P$ does much better than $Q$), while everyone does worse if $u \rightarrow v$ is revealed to everyone! 

This presents an intriguing dilemma for policy makers: should we prefer (i)  a `fair' outcome where everyone suffers equally, or (ii) an outcome which is better for everyone but where some are much better off than others? 
One imagines this scenario might often arise in practice if a large fraction of the population were being guided by a map application provided by one company -- the company might defend a decision to provide faster routing to a select few in order to benefit all. 

Now consider if the privileged few were chosen uniformly at random -- perhaps that might be fair? Notions of fairness beyond equality, and the role of randomness in fairness, were recently explored \cite{random,parity}.

\subsection{Algorithmic Trading}
Algorithmic trading is one area where self-interested agents compete fiercely for high stakes. In many cases, increasing transparency may be beneficial -- but when considering regulation such as the Markets in Financial Instruments Directive (MiFID II), which came into effect in the EU in 2018 and includes various requirements for transparency, one should take care to keep in mind the observation above that in some cases, providing more transparent information to self-interested agents can potentially have negative %lead to adverse 
consequences.

\subsection{Principals and Agents: Actions and Consequences}

In economics, the \emph{principal-agent} problem \cite{Ross73} occurs when one entity (the \emph{agent}) takes actions on behalf of another entity (the \emph{principal}). In general, an agent might act so as to benefit herself even though this could hurt the principal. 
 An example is a fund manager (agent) making investment decisions for an investor (principal). 
The problem is typically worse when the agent has more information than the principal since this makes it hard for the principal to check if the agent is acting in the principal's interest. Hence, it is common to call for greater transparency about the agent.

\cite{Prat05wrong} considers when in fact it might surprisingly be bad for a principal to observe more information about the agent. Prat distinguishes between information directly about the agent's \emph{actions}, and information about the \emph{consequences} thereof. While the latter is helpful, when more information about the actions is available, the agent has an incentive to disregard useful signals which are private to her, and instead to do what an able agent is expected to do a priori. Prat uses this insight to explain a phenomenon previously observed by \cite{Lak92}: US pension funds' performance in equity markets was worse than that of mutual funds. Typically mutual fund investors only see realized investment returns, whereas pension fund investors have much greater access to their fund managers who explain their investment strategy. Thus there may be an incentive for conformism in pension fund managers, leading to lower expected returns.  

In a similar spirit, \cite{Prat05wrong} discusses \emph{executive privilege}, particularly the right of the US President and certain government officers  to resist calls for transparent information about how they arrived at decisions in some settings. % -- \cite{Prat05wrong} 
Prat provides a telling quote from a US Supreme Court ruling relating to the famous Watergate case (US vs. Nixon): ``Human experience teaches us that those who expect public dissemination of their remarks may well temper candor with a concern for appearances and for their own interest to the detriment of the decision-making process.''

\section{Fairness and Discrimination}\label{sec:fair} 

Much work on fairness in machine learning has focused on attempting to avoid discrimination against sub-groups as identified by sensitive features, such as race or gender. Typical metrics for discrimination are based on various types of \emph{disparate impact} or  \emph{disparate treatment} \cite{Bar16}. 
Here we consider a theme which relates to transparency and a common fairness approach used to avoid disparate treatment:  simply remove the sensitive feature(s) from the data. A valid objection to this method is that %in some cases 
it may be possible to predict (and hence reconstruct) the sensitive feature(s) from other features with high confidence. Nevertheless, the approach is in widespread use and the example below may help to explain its intuitive appeal to some.

\begin{figure}[t]
\centering
\includegraphics[width=0.99\linewidth]{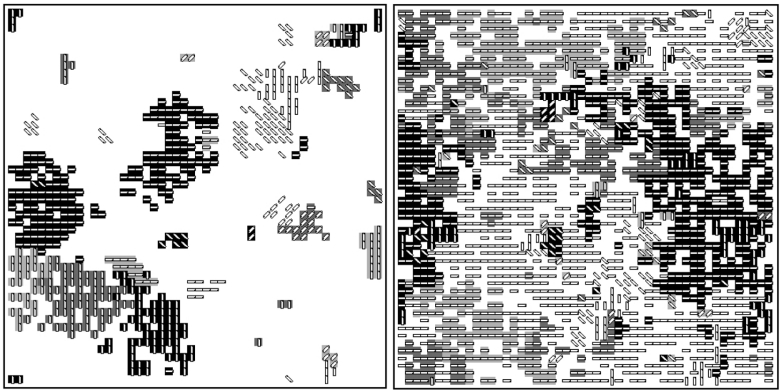} 
\caption{\small Illustrations by \cite{HamAxe06} of the evolution of ethnocentrism. Four different tribes are shown as shades of gray. Agents which are $\{$Ethnocentric (cooperate with their own tribe, defect against other tribes) / Pure cooperators (always cooperate) / Egoists (always defect) / Those who cooperate only with other tribes$\}$ have $\{$Horizontal / Vertical / Diagonal up-right / Diagonal up-left$\}$ lines respectively. The left image shows one run after 100 periods; the right image is after 2000 periods. Over time, almost all agents are ethnocentric. See the reference for a video.}
\label{fig:Axelrod}
\end{figure}

\cite{Axe06} describes fascinating work exploring the conditions under which cooperation naturally emerges in multi-agent populations. In many settings, repeated contact leading to iterated prisoner's dilemma interactions are supportive of the emergence of cooperation %. %\footnote{
(for details of the prisoner's dilemma, see wikipedia or \url{https://plato.stanford.edu/entries/prisoner-dilemma/ }). 

In subsequent work, \cite{HamAxe06} expanded the framework to consider agents which have an extra feature of `ethnicity' which might be regarded as their color or tribe. Now each agent could either cooperate or defect when interacting with other agents of the same tribe; and similarly could either cooperate or defect with agents of other tribes.  Using a simple model of multi-agent evolution, it was strikingly demonstrated that a robust conclusion across a wide range of model parameters, is that `ethnocentrism' emerges. That is, after many iterations, each agent is very likely to cooperate with others of the same tribe, but to defect against agents from other tribes. See Figure \ref{fig:Axelrod}. 

Hence, in this setting, greater transparency (i.e. making faithful information on ethnicity available) leads to  
the emergence of discriminatory behavior. 
In general settings, there is no clear answer as to whether better outcomes will be achieved by hiding or revealing sensitive attributes. Some have attempted to use technical privacy methods to try simultaneously to get the benefits of both \cite{blindjustice}.

Recent work explores multi-agent reinforcement learning 
in repeated social interactions to begin to identify conditions on the environment and the agents' cognitive capacities which lead to the emergence of cooperation (see \cite{Lei17,Low17} and related blog entries).  
\cite{VarVar17} show that bounded rationality with quantized priors may lead to discrimination. 

\section{Trust: Transparency 
and Honesty}\label{sec:trust} 

Many support the view that 
transparency builds trust. For example, the Dalai Lama is reported to have warned ``A lack of transparency results in distrust and a deep sense of insecurity.'' However, we suggest that the story is more nuanced. If we distinguish between transparency -- i.e. the provision of information -- and honesty -- i.e. the accuracy of the information -- then which is more important?

A common view is that 
trust relies on honesty. 
In order to judge trustworthiness, \cite{ONeill13} claims we must examine three qualities: honesty, competence and reliability.   
Stressing honesty, Simon Sinek is reported to have said  ``Trust is built on telling the truth, not telling people what they want to hear.'' 
However, as we can sense from the following phrases themselves, it can be difficult to hear the `harsh truth' from someone speaking with `brutal honesty'.

In recent work, \cite{LevSch15} demonstrated settings where \emph{prosocial lies} increased trust. Prosocial lies are a form of deception -- the transmission of information that intentionally misleads others -- which benefits the target. Their work aims to separate the roles of benevolence and integrity in building interpersonal trust. They conclude that altruistic lies increase trust when deception is directly experienced, or even when it is merely observed.

These effects may make sense when we consider that humans evolved in a tribal multi-agent society. Trusting someone may reflect a belief that they are `in our tribe' 
and will reliably look out for our interests. From a historic survival perspective, this  
benevolence may have been more important than truthful communication -- perhaps a `pre-truth' society?

Interestingly, work in psychology indicates that we are not good at estimating how transparent we are ourselves when communicating with others. \cite{Gil98} showed evidence for the \emph{illusion of transparency} -- a tendency for people to overestimate the extent to which others can discern their internal states. This is attributed to a tendency for people to adjust insufficiently from the `anchor' of their own experience when attempting to adopt another's perspective. Consistent with this, \cite{GrifRoss91} made the following observation: if someone is asked to tap a well-known melody on a tabletop and then to estimate the chance that a listener will be able to identify the song they have tapped, tappers grossly overestimate the listener's abilities.

\section{Machine Interpretability and Understanding}\label{sec:machine}

Human interpretability -- 
that is helping humans to understand machines -- is of great importance. But we describe below 
two classes of `machine interpretability' -- that is helping machines to `understand' -- which are also valuable lines of research.

First, it will be increasingly helpful for machines to be able to follow humans and our motivations. As examples, consider automated care for the elderly, or how an autonomous vehicle waiting at a crossroad should perceive and respond reliably if a human in a car opposite is beckoning to advance. 

Second, we believe that a fruitful line of work will be to help machines  understand each other. Exciting work has begun to explore this direction, looking for ways to enable multiple agents to cooperate effectively \cite{EvtCho17,HavTit17,MorAbb17}. Different paradigms of multi-agent organization should be explored \cite{HorLes04}. One motivation is the goal of AI agents which can autonomously generate and communicate flexible, hierarchical \emph{concepts} which can apply broadly, going beyond traditional transfer learning. Ideally, these concepts will capture high level structure  
in a way which can be transmitted 
efficiently and deployed flexibly. Although it is highly desirable for humans to understand this structure, we suggest that it will still be extremely useful, and perhaps easier, to begin by working on ways for machines to communicate with each other. Further, complex structures may be developed which are beyond easy human understanding. 

It is possible that a relatively low capacity information bottleneck (as examined by \cite{bottle99}) will be a useful constraint between agents to develop such structures. 
 Indeed, \cite{Law17} has suggested that the low inter-human bandwidth of speech compared to the higher processing power of our brains may have helped lead to the development of our own intelligence and consciousness.

With this goal in mind, a useful metric of successful `transfer of understanding' (here a metric for a good explanation) could be to measure how agent B's performance on some task, or range of tasks, improves after receiving some limited information from agent A -- though important details will need to be determined, including suitable bandwidth constraints, and recognition that what is useful will depend on the knowledge that A has already. 

A similar notion might be useful for measuring human interpretability. An example of a helpful conceptual explanation to improve a beginner's ability to play chess might be ``try to control the center.'' One advantage is that only performance improvement  
need be measured, 
bypassing the difficult task of quantifying  
internal understanding directly.

\section{Conclusion}\label{weller:sec:conclusion}

There are many settings where transparency is helpful. We have described some of these settings and have begun to clarify just what sort of transparency may be desirable for each, with accompanying research challenges. This topic is timely given keen interest in laws such as the GDPR (introduced in the EU in 2018) which seek to provide users with some sort of 
meaningful information about algorithmic decisions. 

One focus of this work is to highlight scenarios where transparency may cause harm. We have provided examples where greater transparency can lead to worse outcomes and less fairness.  
We hope to continue to develop frameworks to understand what sorts of transparency are helpful or harmful to whom in particular contexts, and to develop mechanisms which ensure that appropriate benefits are realized. This is a rich area which can draw on connections across economics and the social sciences, philosophy, multi-agent game theory, law, policy and cognitive science.

\section*{Acknowledgements}
This article is an extended version of \cite{orig_chall}. The author thanks Frank Kelly for pointing out the Braess' paradox example and related intuition; and thanks Vasco Carvalho, Stephen Cave, Jon Crowcroft, David Fohrman, Yarin Gal, Adria Gascon, Zoubin Ghahramani, Sanjeev Goyal, Krishna P. Gummadi, Dylan Hadfield-Menell, Bill Janeway, Frank Kelly, Aryeh Kontorovich, Neil Lawrence, Barney Pell and Mark Rowland 
for helpful discussions; and thanks anonymous reviewers for helpful comments. 
The author acknowledges support from the David MacKay Newton research fellowship at Darwin College, The Alan Turing Institute under EPSRC grant EP/N510129/1 \& TU/B/000074, and the Leverhulme Trust via the CFI.

\bibliographystyle{splncs04}
\bibliography{transReferences}
\end{document}